\documentclass[fleqn,twoside]{article}
\usepackage{espcrc2}
\usepackage{epsfig}
\usepackage{graphicx}
\usepackage[figuresright]{rotating}

\newcommand{\AmS}{{\protect\the\textfont2
  A\kern-.1667em\lower.5ex\hbox{M}\kern-.125emS}}
\title{Excited hadrons from improved interpolating fields\thanks{Presented 
at LATTICE 2004 by C.B.\ Lang. 
The work was supported by Fonds zur F\"orderung der 
Wissenschaftlichen Forschung in \"Osterreich (P16824-N08 and
P16310-N08) and by DFG and BMBF.}}

\author{{Tommy Burch$^a$, 
Christof Gattringer$^{a}$, 
Leonid Ya.\ Glozman$^b$, Reinhard Kleindl$^b$,
C.\ B.\ Lang$^b$, 
and Andreas Sch\"afer$^a$
\hspace{1mm}(for the BGR [Bern-Graz-Regensburg] Collaboration)} 
\vskip5mm
$^a$ Institut f{\"u}r Theoretische Physik, Universit{\"a}t
Regensburg, D-93040 Regensburg, Germany. 
\vskip2mm
$^b$ Institut f{\"u}r Physik / Theoretische Physik, Universit{\"a}t
Graz, A-8010 Graz, Austria.
\vskip3mm}
\begin{document}
\begin{abstract}  
The calculation of quark propagators for Ginsparg-Wilson-type Dirac operators
is costly and thus limited to a few different sources.
We present a new approach for determining spatially optimized operators
for lattice spectroscopy of excited hadrons. Jacobi 
smeared quark sources with different widths are combined to 
construct hadron operators with different spatial wave functions.
We study the Roper state and excited $\rho$ and $\pi$ mesons.
\vspace{1pc}
\end{abstract}
\maketitle

\section{Optimized quark sources}

The low-lying hadron spectrum shows a few features which are fingerprints of
QCD. In the meson sector the pion occurs as an  almost-Goldstone boson with its
squared mass vanishing proportional to the quark mass in contrast to all other
mesons. The observed ordering of the lowest positive, $1/2^+, N(1440)$, and
negative parity excitations of the nucleon, $1/2^-, N(1535)$ is 'unnatural'.  A
physical picture based on linear confinement, Coulomb and color-magnetic terms,
always arranges the first radial excitation above the first orbital excitation,
i.e.\ the excited states have alternating parities. 

Whereas ground state spectroscopy on the lattice is by now a well understood
physical problem with impressive agreement with experiment, the lattice study
of excited states is not so far advanced. In a lattice calculation the masses
of excited states show up in the sub-leading exponentials of Euclidean two
point functions. A direct fit of a single correlator is cumbersome since the
signal is  strongly dominated  by the ground state. Also with methods such as
constrained fits  \cite{constrainedfits} or  the maximum entropy method
\cite{maximumentropy} one still needs very high statistics for reliable results
\cite{chenetal,sasaki}.  An alternative method is the variational method
\cite{variation} where one diagonalizes a matrix containing all cross
correlations of a set of several operators with the correct quantum numbers.
For a large enough and properly chosen set of basis operators  each eigenmode
is then dominated by a different physical state.  After normalization the
largest eigenvalue gives the correlator of the ground state, the second-largest
eigenvalue corresponds to the first  excited state, and so on.

It is important to optimize the  spatial properties of the
interpolating operators. An example for this fact  is the Roper
state where the variational method, based on nucleon operators that differ
only in their diquark content but have the same spatial wave function, did
not lead to success \cite{broemmeletal}. It can be argued that a node in
the radial wave function is necessary to  capture reliably the Roper state
or other radially excited hadrons. Recently \cite{prd-paper} we
demonstrated that an elegant solution is to combine Jacobi smeared quark
sources with {\it different} widths to build the hadron operators and
compute the cross-correlations in the variational method. We find good
effective  mass plateaus for the first and partly the second radially
excited states. The propagators are then fitted  using standard
techniques. 

Already in \cite{jacobi1} Jacobi smeared sources were combined with point
sources and cross-correlations studied in similar spirit (see also
\cite{burch}). The technique of Jacobi smearing is well known
\cite{jacobi1,jacobi2}. The  smeared source lives in the timeslice  $t = 0$ 
and is constructed by iterated multiplication with a smearing operator  $H$ on
a point-like source. The operator $H$ is the spatial hopping part of the 
Wilson term at timeslice 0; it  is trivial in Dirac space and acts only on the
color indices. This construction has two  free parameters: The number of
smearing steps $N$ and the hopping parameter $\kappa$. These can be used to
adjust the profile of the source. Here we work with two different sources, a
narrow source $n$ and a wide source $w$ with parameters given by 
\begin{equation}\label{smearparams}
\begin{array}{llll}
n: & 
N = 18\,, &\kappa = 0.210\,, &\sigma /2 \approx 0.27 \; \mbox{fm} \,,\nonumber\\
w: & 
N = 41\,, &\kappa = 0.191\,, &\sigma /2 \approx 0.41 \; \mbox{fm} \,.
\end{array}
\end{equation}
$N$ and $\kappa$ were chosen such that the profiles approximate Gaussian
distributions with the indicated half-widths \cite{prd-paper}. We remark that
the two sources allow the system to build  up radial wave functions with and
without a node. The parameters were chosen such that simple linear 
combinations $c_n \, n \, + \, c_w \, w$  of the narrow and wide profile
approximate the first and second radial wave functions of the spherical
harmonic oscillator: The coefficients $c_n \sim 0.6, c_w \sim 0.4$ approximate
a  Gaussian with a half-width of $\sigma/2 \sim 0.33$ fm, while $c_n \sim 2.2,
c_w \sim -1.2$ approximate the corresponding excited wave function with one
node. 

The final form of the wave function  is determined through the variational
method \cite{variation}. In this approach one computes a complete correlation
matrix of operators $O_i, i = 1,2, \, ... \, R$  that create from the vacuum
the state which one wants to analyze. The eigenvalues  $\lambda^{(k)}(t)$ of
the correlation matrix behave as $ \lambda^{(k)}(t) \propto e^{-t \, M_k} [1 +
{\cal O}(e^{-t \, \Delta M_k})]$,  where $\Delta M_k$ is the distance of $M_k$
to nearby energy levels. The hadron sources we use for the correlation matrix
are constructed from the  narrow and wide quark sources. 

\section{Excited nucleon signals}

For our quenched calculation we use the chirally improved Dirac operator
\cite{chirimp}. It is an approximation of a solution of the Ginsparg-Wilson
equation  which governs chiral symmetry on the lattice. This operator is well
tested in quenched ground state spectroscopy \cite{bgr}  where pion masses down
to 250 MeV can be reached at a considerably smaller  numerical cost than needed
for exact  Ginsparg-Wilson fermions. For ground states the chirally improved
action shows very good scaling behavior.  The gauge configurations were
generated on a  $12^3 \times 24$ lattice with the L\"uscher-Weisz action
\cite{Luweact}. The inverse gauge coupling is $\beta = 7.9$, giving rise to a
lattice spacing of $a = 0.148(2)$ fm as determined from the Sommer parameter
\cite{scale}. The statistics of our ensemble is 100 configurations. We use 10
different quark masses $m$ ranging from $am = 0.02$ to  $am = 0.20$.

Our analysis is based on the interpolator $\varepsilon_{abc} (u_a C \gamma_5
d_b)u_c$.  Each of the three quarks can be smeared either narrow ($n$) or wide
($w$). This gives 8 possible combinations ($nnn$, $nnw$, etc.). From a subset
of 4 of these operators (after projection to definite parity) we calculate the
correlation matrix $C(t)$ which we then use in the variational method.  The
exponential decay of three eigenvalues is clearly identified.  We identify
these signals with the nucleon, the Roper state and  the next positive parity
resonance $N(1710)$. A detailed discussion of further checks on the correct
identification of the Roper state can be found in \cite{prd-paper} (see also
\cite{chenetal} concerning the problem of nucleon-$\eta^\prime$ ghost 
contributions).

\section{Excited meson signals}

\begin{figure}[t]
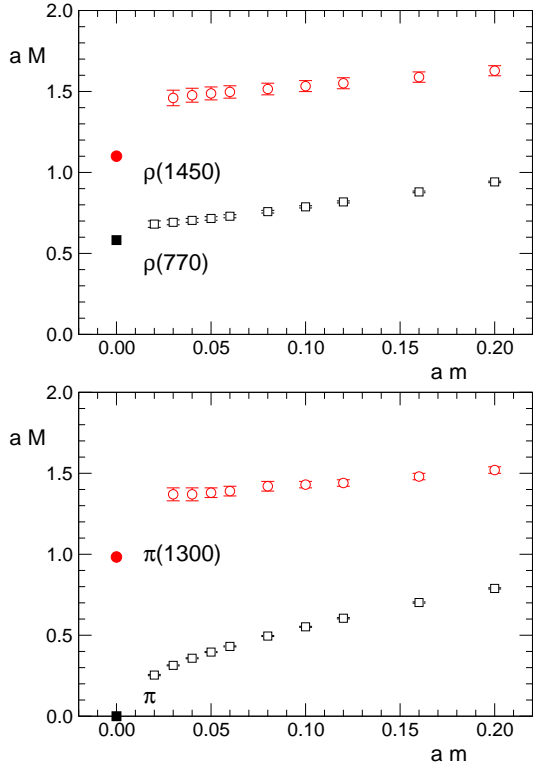

\begin{center}
\includegraphics*[width=7cm]{m_rho_plot.eps}\\
\includegraphics*[width=7cm]{m_pi_plot.eps}
\end{center}
\vspace*{-10mm}
\caption{Masses of $\rho(770)$, $\rho(1450)$, $\pi(140)$ and $\pi(1300)$ 
as a function of the quark mass (the experimental data were converted to lattice units with
the Sommer parameter scale). 
\label{fig2}}
\vspace*{-2mm}
\end{figure}

As another test of our approach we discuss the $\pi$- and $\rho$-mesons and
their radial excitations. For the $\rho$ we use the interpolators 
$\overline{u}(x) \, \gamma_i \, d(x)$ and  $\overline{u}(x) \, \gamma_4
\,\gamma_i \, d(x)$, for the pion  $\overline{u}(x) \, \gamma_5 \, d(x)$ and 
$\overline{u}(x) \, \gamma_4 \,\gamma_5 \, d(x)$. Again we use wide and narrow
quark sources for both interpolating fields, corresponding to 3 operators each
(the combinations $nw$ and $wn$ give identical correlators and one of them can
be omitted).  When diagonalizing the $3 \times 3$ matrix with either
interpolator  we see a pronounced exponential decay only for the two larger (in
magnitude) eigenvalues, $\lambda^{(1)}(t)$ and $\lambda^{(2)}(t)$. The smallest
eigenvalue $\lambda^{(3)}(t)$  does not show a clear effective mass plateau.
This is an indication that this eigenvalue couples to an unphysical quenched
ghost state \cite{Bardeen,DeGrand,chenetal}.  The final results for the masses
as a function  of the quark mass are shown in Fig.\ \ref{fig2}. 

We find that the ground state meson masses approach their experimental values
reasonably well. The excited state masses are considerably above their
experimental values. There is, however, a plausible reason for this behavior.
The sizes of hadrons which are not, or only weakly, affected by spontaneous
chiral symmetry breaking can be estimated from the known string tension,  which
is approximately 1 GeV/fm. Hence the size of the excited mesons should be
larger than the ground state, about 1.5 fm. Thus the size of our lattice (1.8
fm) is clearly not enough for a precise measurement of e.g.\ the $\rho(1450)$
mass. The finite size effect cannot be neglected for the excited state since it
apparently shifts the measured mass up as compared to the experimental value. 

A crucial test of our method is to check whether indeed the ground state is built from a
nodeless combination of our  sources and the excited states do show nodes. 
This question can be addressed by analyzing the eigenvectors of the 
correlation matrix. This has been done in \cite{prd-paper} and indeed confirms
the expectation.

\end{document}